\documentclass[%
aip,
 amsmath,amssymb,
 reprint,
 superscriptaddress,
 floatfix,
]{revtex4-1}

\usepackage{graphicx}
\usepackage{dcolumn}
\usepackage{bm}
\usepackage[unicode,colorlinks=true,allcolors=blue]{hyperref} 
\usepackage{siunitx} 
\usepackage{cleveref} 
\usepackage[T1]{fontenc} 
\usepackage[version=4]{mhchem}  
\usepackage[dvipsnames]{xcolor}
\usepackage[normalem]{ulem}

\usepackage{mathtools}

\usepackage[acronym]{glossaries}
\newacronym{acf}{ACF}{autocorrelation function}
\newacronym{bnmr}{$\beta$-NMR}{$\beta$-detected nuclear magnetic resonance}
\newacronym{md}{MD}{molecular dynamics simulation}
\newacronym{nmr}{NMR}{nuclear magnetic resonance}
\newacronym{ps}{PS}{polystyrene}
\newacronym{vft}{VFT}{Vogel-Fulcher-Tammann}
\glsdisablehyper

\newcommand*{\revII}[1]{\textcolor{black}{#1}}
\newcommand*{\revI}[1]{\textcolor{black}{#1}}
\newcommand*{\revIdel}[1]{}   

\begin{document}

\preprint{AAPM/123-QED}

\title{Energy barriers and cooperative motion at the surface of freestanding glassy polystyrene films}

\author{D.~Fujimoto}
\email{fujimoto@phas.ubc.ca}
\affiliation{Department of Physics and Astronomy, University of British Columbia, Vancouver, BC V6T~1Z1, Canada}
\affiliation{Stewart Blusson Quantum Matter Institute, University of British Columbia, Vancouver, BC V6T~1Z4, Canada}

\author{W.~A.~MacFarlane}
\affiliation{Stewart Blusson Quantum Matter Institute, University of British Columbia, Vancouver, BC V6T~1Z4, Canada}
\affiliation{Department of Chemistry, University of British Columbia, Vancouver, BC V6T~1Z1, Canada}
\affiliation{TRIUMF, Vancouver, BC V6T~2A3, Canada}

\author{J.~Rottler}
\email{jrottler@physics.ubc.ca}
\affiliation{Department of Physics and Astronomy, University of British Columbia, Vancouver, BC V6T~1Z1, Canada}
\affiliation{Stewart Blusson Quantum Matter Institute, University of British Columbia, Vancouver, BC V6T~1Z4, Canada}

\date{\today}

\begin{abstract}
    We investigate the near-surface relaxation of freestanding atactic \glsdesc{ps} films with molecular dynamics simulations. As in previous coarse-grained simulations, relaxation times for backbone segments and phenyl rings are linked to their bulk relaxation times via a power law coupling relation. Variation of the coupling exponent with distance from the surface is consistent with depth-dependent activation barriers. We also quantify a reduction of dynamical heterogeneity at the interface which can be interpreted in the framework of cooperative models for glassy dynamics.  
\end{abstract}
\maketitle 
\glsresetall	

\section{Introduction}
Polymers have a high degree of mechanical and chemical tunability, making them extremely versatile materials. Upon cooling, many polymers vitrify rather than crystallize. Long before a crystal forms, molecular motion becomes frozen and dynamical timescales quickly surpass those accessible in experiments. Boundary conditions have a strong impact on these relaxation dynamics in glassy polymers \cite{Ediger2014}. In freestanding films with a vacuum interface, the glass transition temperature, $T_\mathrm{g}$, is reduced as the thickness decreases \cite{forrest_glass_2001,sharp_free_2003}. Moreover, the layer-resolved segmental (relaxation) dynamics accelerates substantially as the depth $z$ below the free interface decreases \cite{peter_thickness_2006,paeng_jacs_2011,Mckenzie2018,Mckenzie2015}. Recent evidence from \gls{md} \cite{diaz2018temperature} and theoretical arguments \cite{schweizer2019progress} suggest that the molecular relaxation time $\tau(z,T)$ near the surface is coupled to the bulk relaxation time $\tau_\mathrm{\revII{bulk}}(T)$ via a power-law relation 
\begin{equation}    \label{eq:coupling-prop}
    \tau(z,T)\sim \tau_\mathrm{\revII{bulk}}(T)^{f(z)},
\end{equation}
with a ``coupling exponent'' $f(z) \in [0,1]$ capturing the dependence on depth $z$. The origin of this power-law form, as explained below, lies in the exponential dependence of the relaxation time on an energetic barrier for activated motion. 

Several theoretical pictures have been proposed that arrive at the same functional form  \cref{eq:coupling-prop}, but differ in their interpretation of the coupling exponent.  The ``elastically collective nonlinear Langevin equation'' (ECNLE) theory of Schweizer and co-workers\cite{mirigian_communication_2014}, for instance, proposes a reduction of the activation barrier via modified local caging constraints due to loss of neighbors as well as truncation of long range elastic interactions\cite{phan2018dynamic,phan2019influence,phan2019theory,phan2020theory}. Another picture asserts that the coupling exponent reflects a temperature and distance-dependent reduction of the size of string-like cooperative mobile regions as the major driver of interfacial relaxation \cite{salez_cooperative_2015}. A recent study by \citet{zhang2019collective}, however, indicates that the length of such mobile strings varies only weakly near the interface, although the dynamical scale of this layer \cite{lang_interfacial_2013} is proportional to the length of mobile strings \cite{hanakata_interfacial_2014,shavit_physical_2014}. Gaps thus remain in our understanding of interfacial dynamics of glass-forming materials.

The present study presents \gls{md} simulations of freestanding atactic \gls{ps} films at the united-atom level. It builds on previous results of \citet{Zhou2017}, who computed the layer-resolved segmental relaxation times in \gls{ps}-films of up to \SI{28}{\nm} thickness by monitoring the angular displacement along the polymer backbone (see \Cref{fig:setup}). Here we focus additionally on the rotational dynamics of the phenyl sidegroups that reflect (slightly faster) $\gamma-$relaxation processes \cite{Vorselaars2007}. The phenyl ring motion is particularly important for the interpretation of \gls{bnmr} experiments, because the ${\rm Li}^+$-ions are expected to be bound between such rings \cite{Mckenzie2014,Mckenzie2015}. We determine the coupling exponent that describes the dynamics at the surface, and show that its functional form is consistent with an average activation barrier that varies with depth. We also compute, as one measure of cooperativity, the dynamical four-point susceptibility $\chi_4(T,z,t)$, and find that it decreases strongly at the surface. A coupling exponent based on this parameter can therefore also describe the observed variation of relaxation times with depth.

\section{Descriptions of interfacial dynamics}

\subsection{Distance dependent energy barrier} \label{sec:energy}

In polymeric glass formers, the temperature dependence of the bulk relaxation time typically exhibits thermally activated behavior, which is well-described over some range of $T$ by the \gls{vft} equation. In a film, it is reasonable to expect that the barrier for activated motion, as well as the exponential prefactor, become explicitly dependent on the depth $z$, such that the \gls{vft} equation reads
\begin{equation}    \label{eq:tau_vft}
    \tau(z,T)=\tau_0(z)\exp\left[\frac{\Delta E(z)}{k_B(T-T_0)}\right],
\end{equation}
where $T_0$ denotes the Vogel temperature, and the effective activation energy barrier $\Delta E(z)$ reflects an average over a distribution of local energy barriers for molecular motion. \revI{To avoid overparameterizing the model, we assume a depth-independent $T_0$ and show below that such a model provides an excellent fit to the simulation data.} The bulk relaxation time is $\tau_\mathrm{\revII{bulk}}(T) = \tau(\infty,T)$. After dividing by $\tau_\mathrm{\revII{bulk}}(T)$, equation (\ref{eq:tau_vft}) can be rearranged as
\begin{equation}    \label{eq:coupling_exp1}
    \frac{\ln(\tau(z,T)/\tau_0(z))}{\ln(\tau_\mathrm{\revII{bulk}}(T)/\tau_0)}=\frac{\Delta E(z)}{\Delta E_\infty}=f(z),
\end{equation}
or alternatively
\begin{equation}    \label{eq:coupling1}
    \frac{\tau(T,z)}{\tau_0(z)}=\left(\frac{\tau_\mathrm{bulk}(T)}{\tau_0}\right)^{f(z)}.
\end{equation}
This simple heuristic derivation yields a coupling relation between bulk and surface dynamics with a temperature-independent coupling exponent as introduced by \citet{diaz2018temperature}. It can be expected to hold below an onset temperature where $\tau_\mathrm{\revII{bulk}}(T)\gg \tau_0\equiv \tau_0(\infty)$ and the interfacial dynamics ``decouples'' from the bulk and becomes faster. At higher temperatures, however, the coupling exponent $f(z)\simeq 1$ and the interfacial dynamics is strongly coupled to the bulk. Simulations for vacuum interfaces suggest that below the onset temperature the coupling exponent is temperature-independent and has an exponential depth-dependence, $f(z)=1-\epsilon_0\exp(-z/\xi_{\Delta E})$ with $\xi_{\Delta E}$ an interfacial length scale \cite{diaz2018temperature}.

\subsection{ECNLE theory} \label{sec:ECNLE}
\revI{In the above empirical treatment, the energy barrier factorization $\Delta E(z)=\Delta E_{\infty} f(z)$ is in some sense a consequence of the assumption of a $z$-independent Vogel temperature. The factorization property and the resultant (de)coupling relation can be justified with much more rigor in the microscopic ECNLE theory \cite{phan2018dynamic,phan2019influence,phan2019theory,phan2020theory}. Here, the central object is a dynamic free energy barrier $F_\mathrm{dyn}(T,r,z)$, where $r$ denotes the displacement from a particle or segment from a local equilibrium position. Several physical mechanisms are included to capture the influence of an interface on this barrier: (i) loss of nearest neighbors immediately at the surface, (ii) a transfer mechanism by which the less constrained surface particles in turn provide fewer caging constraints in the layers below and (iii) a modification of collective, long ranged elastic contributions via lowering and truncation of displacement field amplitudes \cite{schweizer2019progress}. Based on this physical picture, the theory is able to predict a factorization of the dynamical free energy into separate dependencies on distance and temperature/density, \cite{phan2019influence}}
\begin{equation}
    F_{\rm total}^{\rm film}(T,z)\approx F_{\rm total}^{\rm bulk}(T)f(z).
\end{equation}
\revI{If one furthermore assumes activated dynamics for barrier crossing, $\tau(z,T)\sim \exp\left[F_{\rm total}^{\rm film}(T,z)/k_BT\right]$, a coupling relationship of the type of equation (\ref{eq:coupling-prop}) immediately follows. Moreover, as a result of mechanism (ii) mentioned above, the coupling exponent is predicted to have an exponential depth dependence with a short characteristic length of \num{\sim 1.4} particle diameters. All of the temperature dependence is carried by the behavior of the bulk material. As a result, the coupling exponent can ultimately be related to the gradient of the glass transition temperature. The theory has recently been extended beyond a description of the mean relaxation time by including barrier fluctuations via Gaussian distributions of local volume fractions \cite{xie2020collective,xie2020microscopic}. }

\subsection{Cooperative strings} \label{sec:strings}

An alternative approach posits that the origin of the enhanced surface relaxation is a reduction of the number of particles involved collectively in a structural relaxation event. It builds on the observation (mainly from simulations) that mobile particles in glasses organize themselves in a string-like form such that $N^*(T)$ particles have to relax for one particle to escape from a local cage. \citet{salez_cooperative_2015} start from a free volume picture and write the probability for an $N$-particle relaxation process along a string in the bulk:
\begin{equation}    \label{eq:prob_relax_n}
    P_N(T) \sim \frac{1}{\lambda^3\tau_c} \epsilon^{N-1} (1-\epsilon)\theta(N-N^*(T)),
\end{equation}
where $\tau_c$ is an `onset' timescale, $\lambda$ an average intermolecular distance, and $\epsilon=\tau_0/\tau_c\ll 1$ is an elementary `coherence probability'. Since particles in a cooperative string need to move in phase, one expects the probability to decrease exponentially with the string length. The total probability for relaxation,
\begin{equation}    \label{eq:prob_relax_total}
    P(T)=\sum_{N=N^*}^\infty P_N(T) \sim \frac{1}{\lambda^3\tau_c}\epsilon^{N^*-1},
\end{equation}
is dominated by the threshold string length $N^*$. Defining the bulk relaxation time as $\tau_\mathrm{\revII{bulk}}(T)\sim 1/P(T)$, one obtains 
\begin{equation}    \label{eq:coupling2}
    \frac{\tau_\mathrm{\revII{bulk}}(T)}{\tau_0}\sim\left(\frac{\tau_c}{\tau_0}\right)^{N^*(T)}.
\end{equation}
\citet{salez_cooperative_2015} now generalize this expression for the bulk relaxation time to free interfaces by replacing $N^*(T)$ with $N^*(z,T)=N^*(T)f(z/\xi_\mathrm{\revII{bulk}}(T))$ where $f(z/\xi_\mathrm{\revII{bulk}}(T))\le 1$ reflects a reduction of the length of the cooperative string near the surface. This reduction can be expected to occur over a scale set by the temperature-dependent bulk cooperative length scale $\xi_\mathrm{\revII{bulk}}(T)$. Interestingly, this yields a formula for the relaxation times near free interfaces that has the same form as \cref{eq:coupling1}, 

\begin{equation}    \label{eq:string-eq}
    \frac{\tau(T,z)}{\tau_0}=\left(\frac{\tau_\mathrm{\revII{bulk}}(T)}{\tau_0}\right)^{f(z,T)},
\end{equation}
but the coupling exponent now reads
\begin{equation}    \label{eq:coupling_exp2}
    f(z,T)=\frac{N^*(z,T)}{N^*(T)}
\end{equation} 
and depends explicitly on temperature. Moreover, $\tau_0$ is assumed to be independent of position and just reflects a microscopic timescale. The fact that the same functional form arises in two seemingly independent derivations can be traced to the exponential dependence of the probability for relaxation on the size of the cooperative region. The model thus embodies the central tenet of the Adam-Gibbs argument, namely that the activation barrier is proportional to the number of particles in the cooperatively rearranging region. The string model of (bulk) glassy dynamics \cite{betancourt2014} makes this explicit, 
\begin{equation}    \label{eq:tau_starr}
\tau(T)\propto \big[\exp(\Delta\mu/k_BT)\big]^{f(T)}
\end{equation}
with $f(T)=L(T)/L(T_A)$, where $L(T)$ is the length of the cooperative string and $\Delta \mu$ is the activation barrier at an onset temperature $T_A$. \revI{The relevance of these string-like excitations has, however, been called into question by a recent computer simulation study \cite{hung2020}.}
\begin{figure}[t]
\includegraphics[width=0.8\columnwidth]{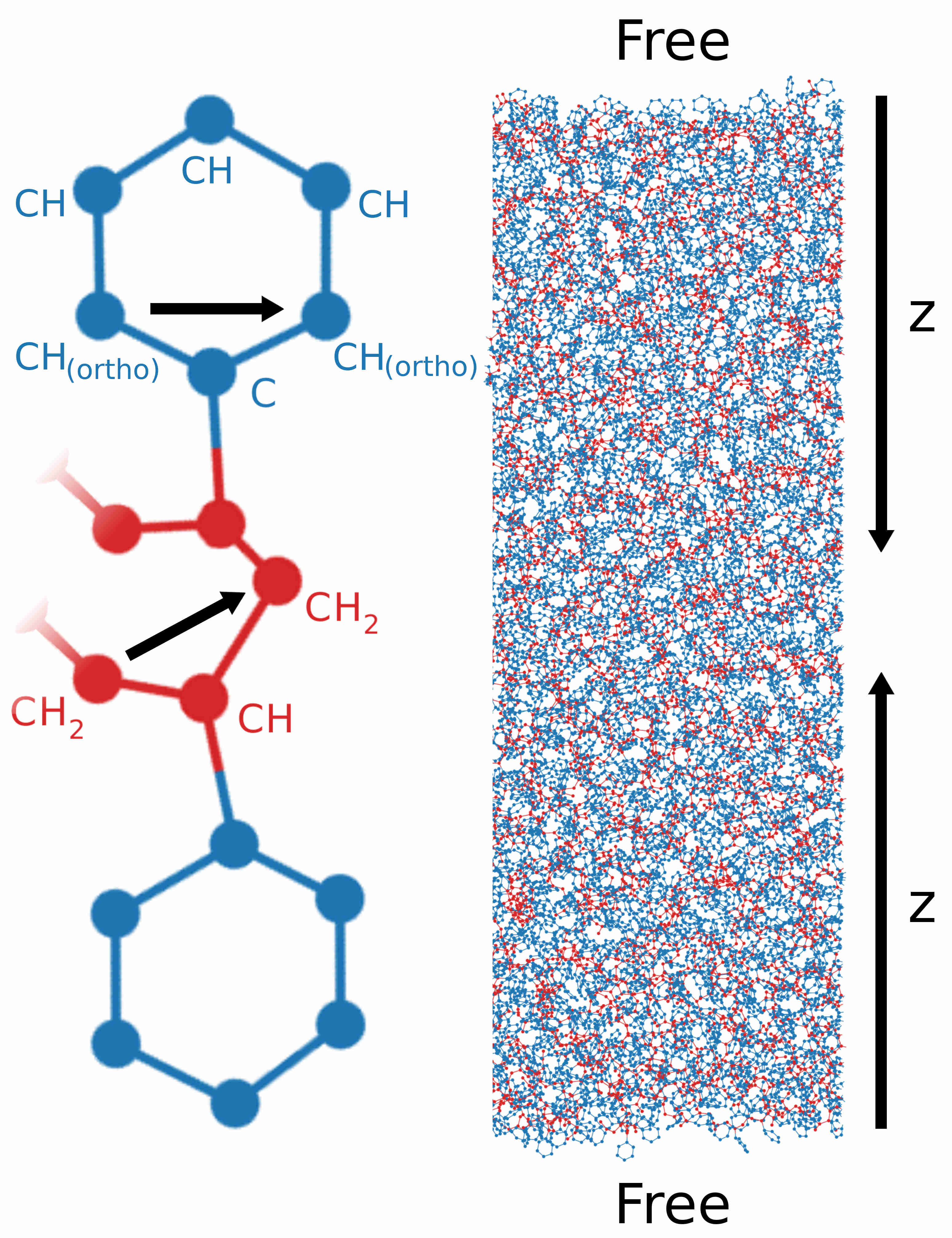}
\caption{Snapshot of a simulated free-standing atactic \glsdesc{ps} film at $T=\SI{200}{\K}$. The distance between the two surfaces is about \SI{31}{\nm}. Vectors connecting ortho-atoms and adjacent \ce{CH2} united atoms on the backbone are used to measure the polymer dynamics.} 
\label{fig:setup}
\end{figure}

\section{Simulation methods} 

A united atom model of atactic \gls{ps} introduced previously by \citet{Vorselaars2007} was used to simulate free standing films \SI{\sim30}{\nm} thick. The \num{\sim4e4} atom simulation was composed of 500 polymer chains, each 10 monomer units in length. Molecular dynamics simulations were carried out using the LAMMPS package \cite{Plimpton1995} in an NVT ensemble with a Nos\'e-Hoover thermostat. The equations of motion were integrated with a time step of \SI{2}{\fs} in a velocity-Verlet scheme. Periodic boundaries were used along both $\hat{x}$ and $\hat{y}$, and reflective walls were used along $\hat{z}$, with final box dimensions fixed to $5.5\times 5.5\times \SI{40}{\nm}$. To prevent drift, the center of mass linear momentum was re-scaled to zero at every time step.  
	
\begin{figure}[t]
\includegraphics[width=\columnwidth]{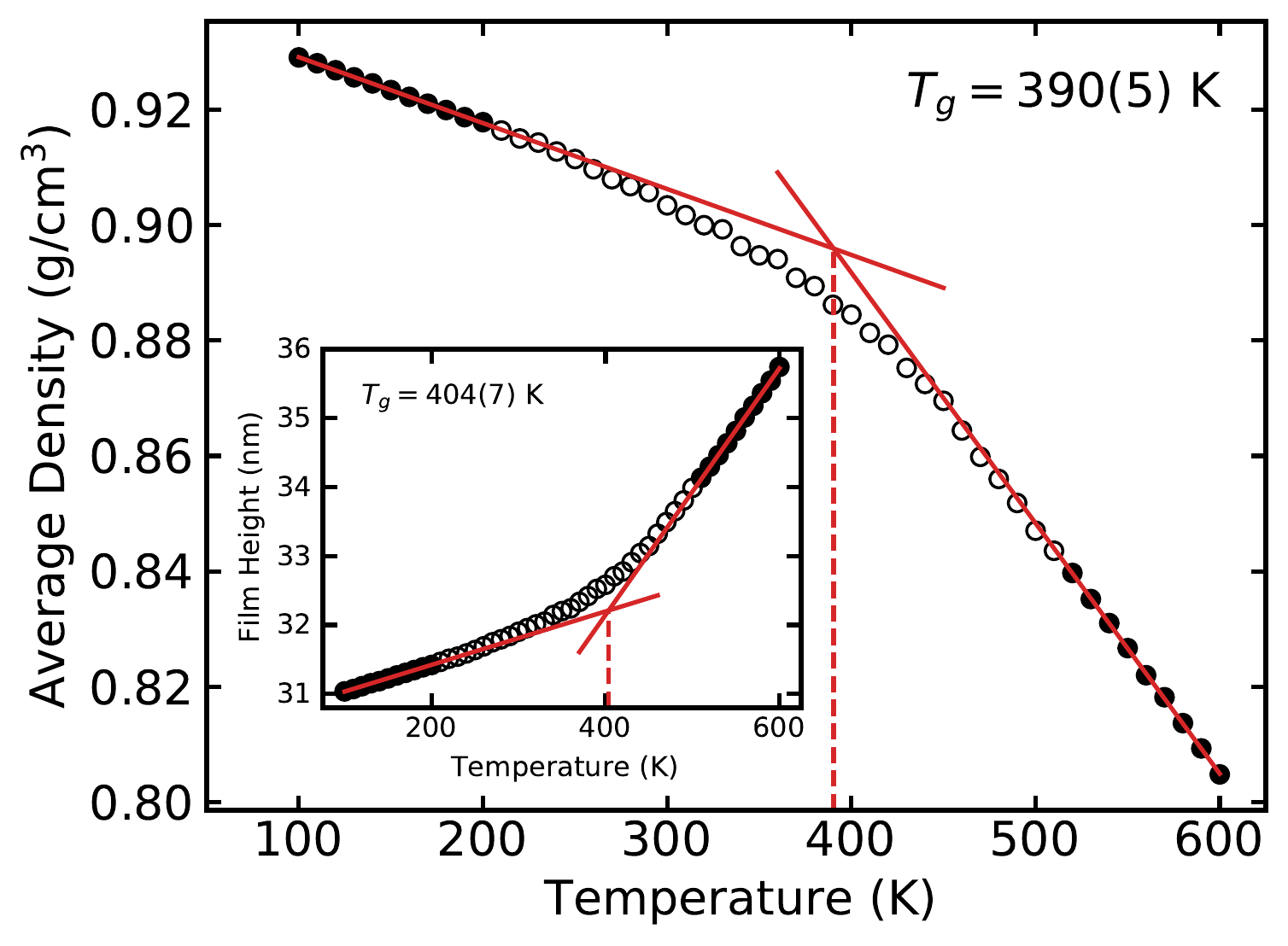}
\caption{\revII{Average film density and film height (inset)} as a function of temperature during cooling of the \glstext{ps}-film. The glass transition temperature, $T_\mathrm{g}$ was found by fitting the linear regions (fitted points indicated by the filled symbols). \revII{Film edges were defined by the Gibbs dividing surface\cite{Hudzinskyy2011a}.} }
\label{fig:tg}
\end{figure}

The film was generated by placing the polymer chains in a $40\times40\times\SI{40}{\nm}$ simulation box and equilibrating at \SI{600}{\K} for \SI{5}{\ns}, accommodating for placement overlap by limiting atomic motion to \SI{0.1}{\angstrom} for the first \SI{10}{\ps}. The box was then compressed to \revII{a cube of side length \SI{5.5}{\nm}} over \SI{10}{\ns}. After another \SI{5}{\ns}, the reflective walls \revII{along $\hat{z}$} were relaxed to their initial positions over the course of \SI{10}{\ns}, and an additional \SI{5}{\ns} was allowed to pass. The film was then cooled at \SI{0.01}{\K\per\ps}, which is a typical rate used in  \gls{md}\cite{Zhou2017,Vorselaars2007}. The glass transition temperature $T_\mathrm{g}$ of the film was found to be \revII{\SI{390\pm5}{\K}} using the \revII{average film} density, and \revII{\SI{404\pm7}{\K}} using the film height (\Cref{fig:tg}), where the film edge was defined \revII{by the Gibbs dividing surface as illustrated by \citet{Hudzinskyy2011a}.} These values are within a few degrees from those reported by Zhou and Milner for a \gls{ps} film also composed of 10mers of comparable thickness \cite{Zhou2017}. From \SIrange{600}{100}{\K}, the film height decreased from \revII{\SIrange{35.7}{31.0}{\nm}}, and the density increased from \revII{\SIrange{0.80}{0.93}{\g\per\cm^3}}. 

The motion of the two local structure vectors ${\vec v}(t)$ connecting the ortho atoms in the phenyl rings (adjacent to the tethering bond between the ring and the backbone) as well as adjacent \ce{CH2} united atoms on the backbone were considered as indicators of polymer dynamics, as depicted in \Cref{fig:setup}. The \gls{acf} of the second Legendre polynomial of the normalized vectors, 
\begin{equation}\label{eq:autocorr}
C(t) = \frac{3}{2}\Big\langle \big[\hat{v}(t')\cdot\hat{v}(t'-t)\big]^2 \Big\rangle_{t'}-\frac{1}{2},
\end{equation}
was used to determine the relaxation time, $\tau(z,T)$, defined to be time it takes for the average \gls{acf} to decay by a factor of $1/e$. The \gls{acf} was averaged by grouping each vector into \SI{1}{\nm} thick parallel laminae by distance to the nearest free surface.

\section{Results}

\begin{figure}[t]
    \centering
    \includegraphics[width=8cm]{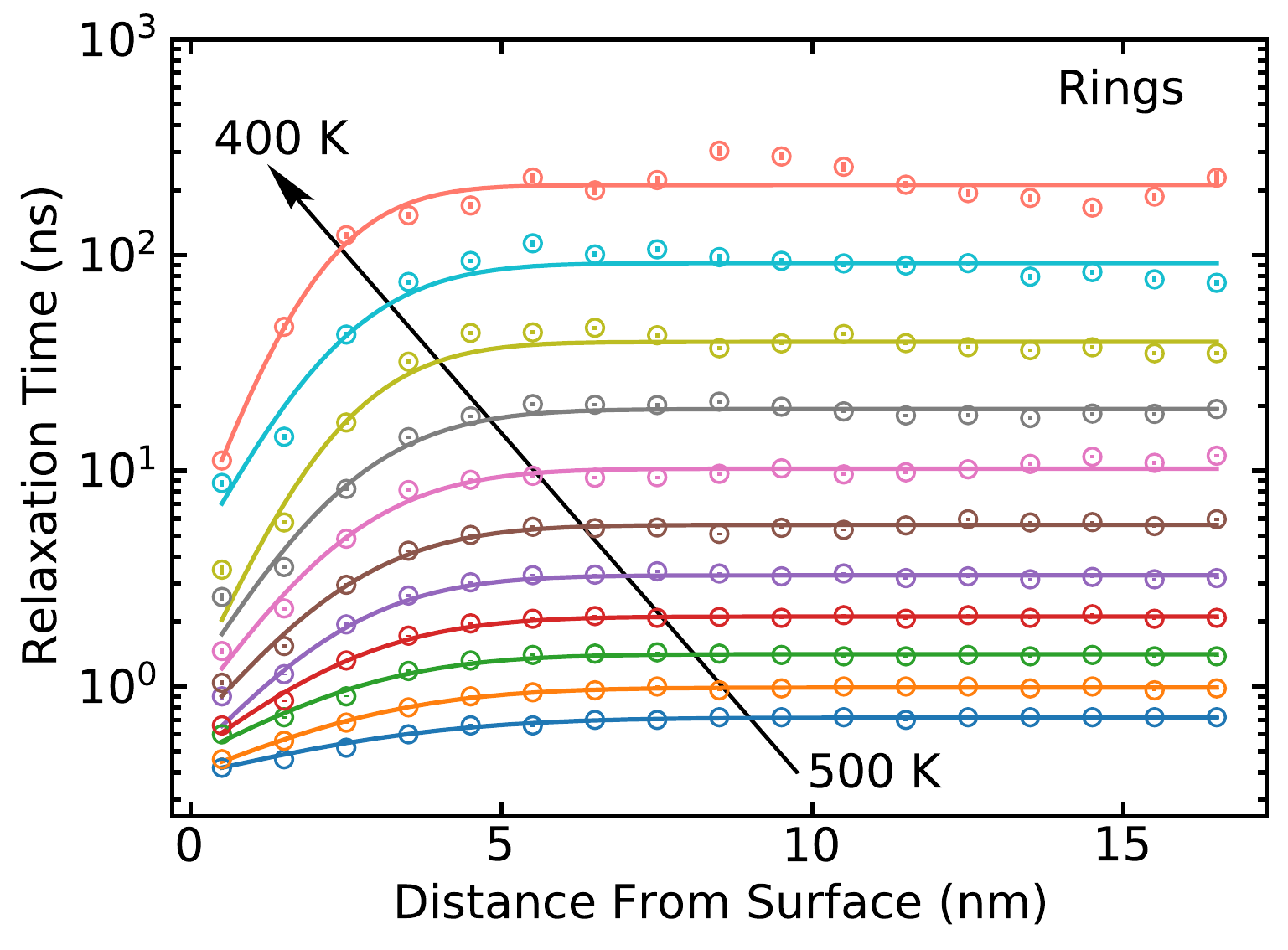}
    \includegraphics[width=8cm]{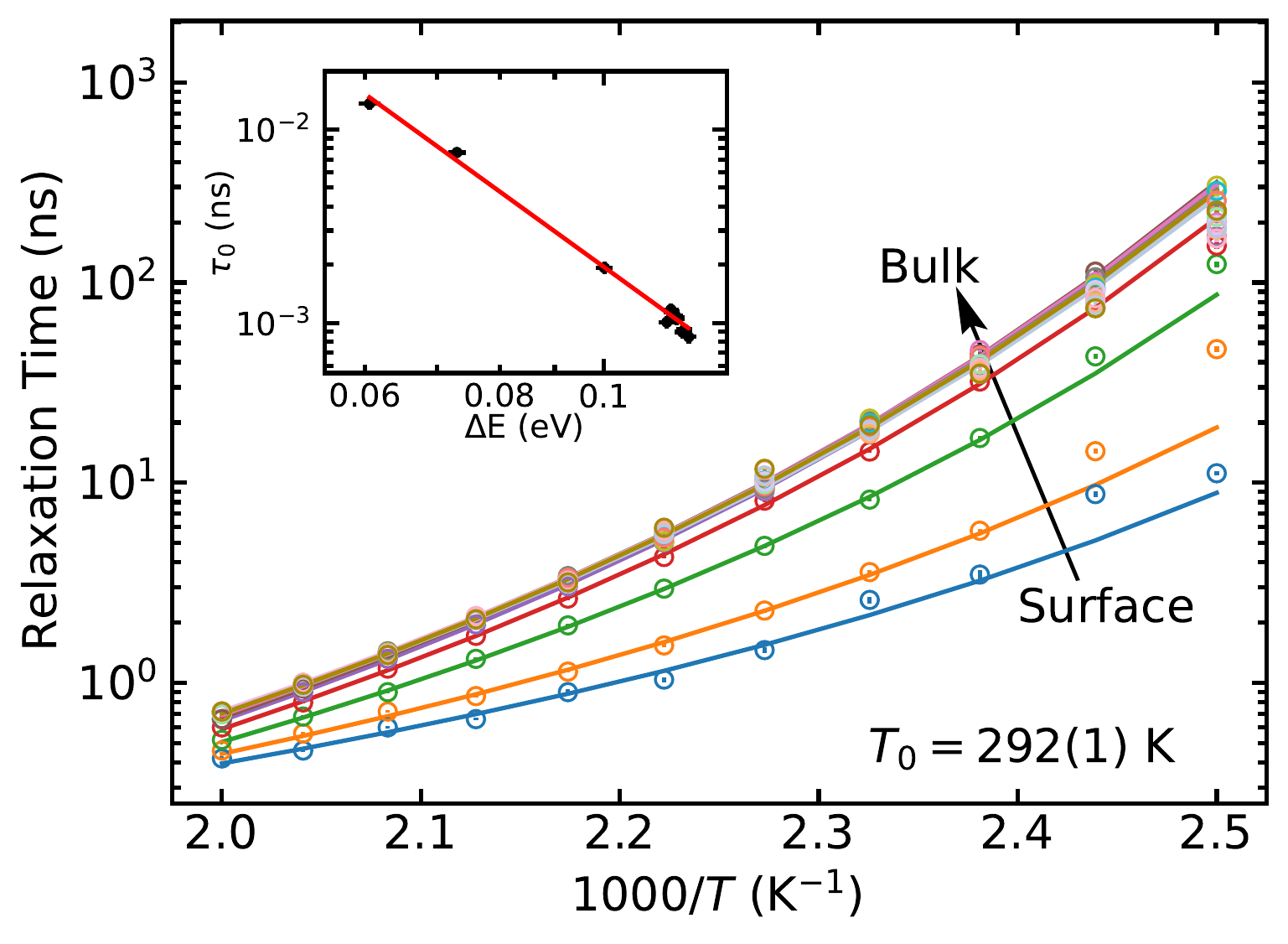}
    \caption{Distance from surface \emph{(top)} and temperature \emph{(bottom)} dependence of the time of the \glsdesc{acf} given by \cref{eq:autocorr} to decay to $1/e$, corresponding to the rotational motion of the \glsdesc{ps} phenyl rings. Also shown are \gls{vft} fits to \cref{eq:tau_vft} with a global \gls{vft}-temperature $T_0=\SI{292\pm1}{\K}$. The inset shows that the preexponential factor and activation barrier follow the Meyer-Neldel rule. }
    \label{fig:depth_vft}
\end{figure}

\begin{figure}[t]
    \centering
    \includegraphics[width=8cm]{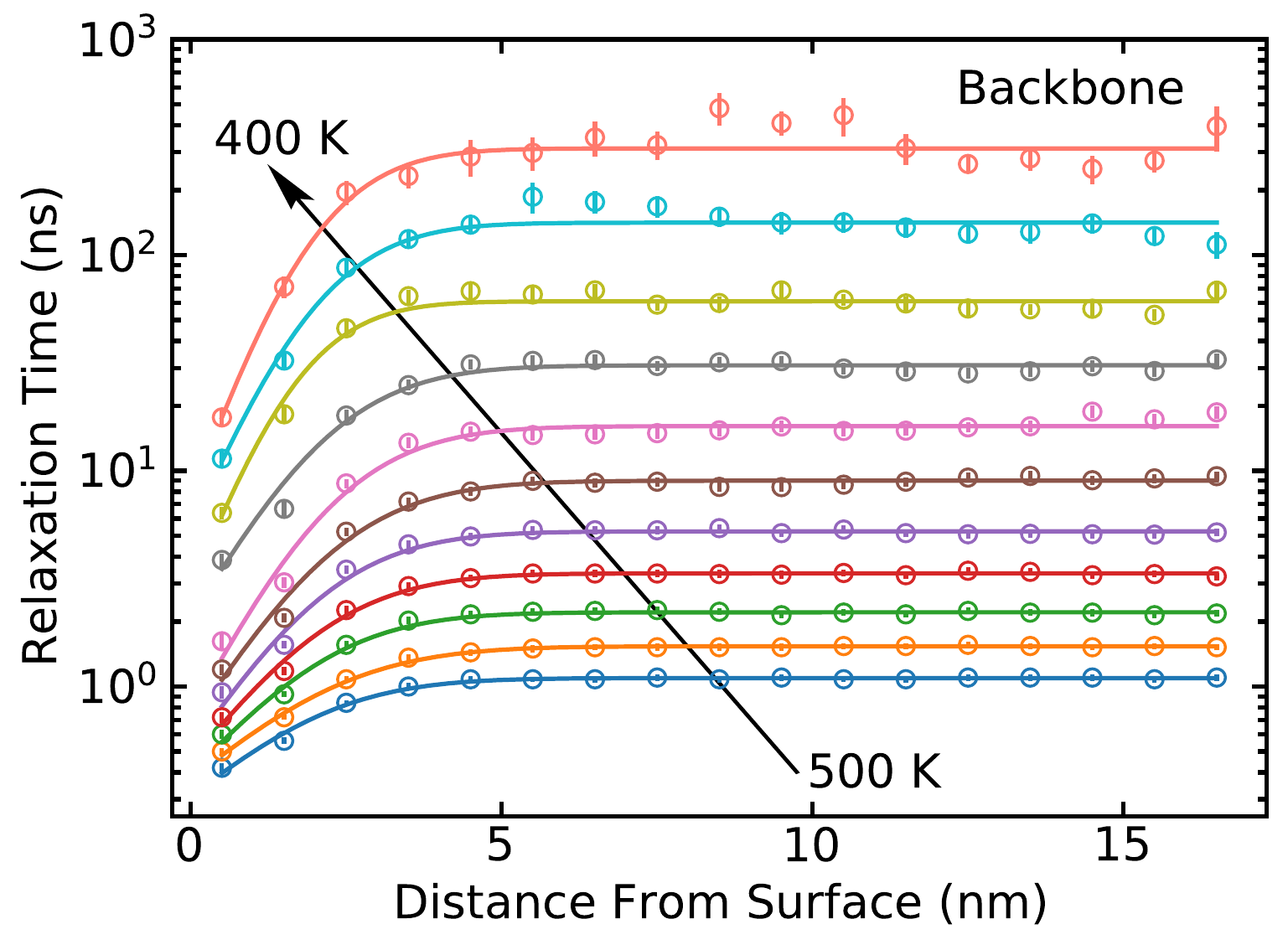}
    \includegraphics[width=8cm]{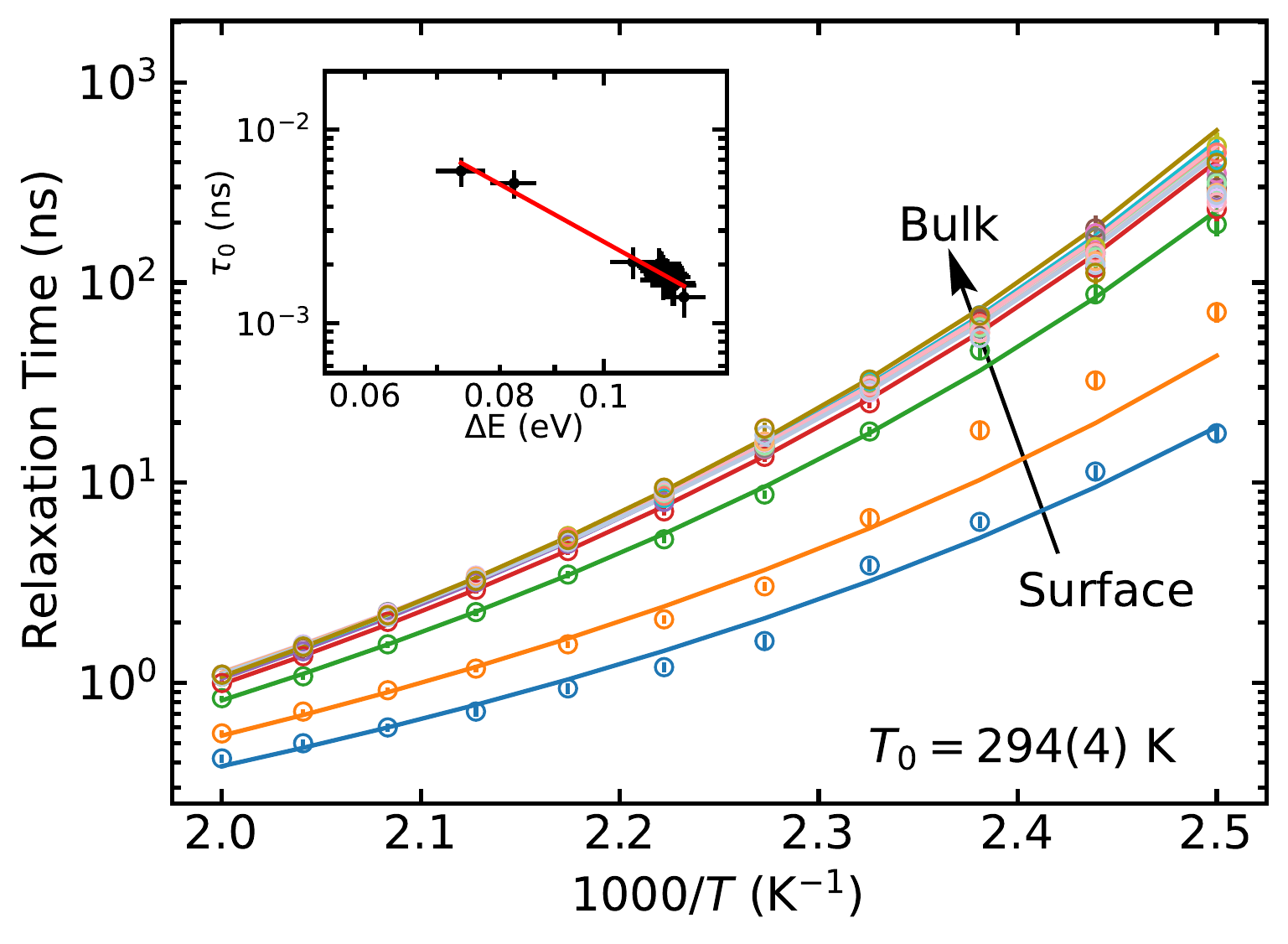}
    \caption{Distance from surface \emph{(top)} and temperature \emph{(bottom)} dependence of the time of the \glsdesc{acf} given by \cref{eq:autocorr} to decay to $1/e$, corresponding to the relaxation of the \glsdesc{ps} backbone segments. Also shown are \gls{vft} fits to \cref{eq:tau_vft} with a global \gls{vft}-temperature $T_0=\SI{294\pm4}{\K}$. The inset shows that the preexponential factor and activation barrier follow the Meyer-Neldel rule. }
    \label{fig:depth_vft_bb}
\end{figure}

\begin{figure}[t]
\includegraphics[width=0.9\columnwidth]{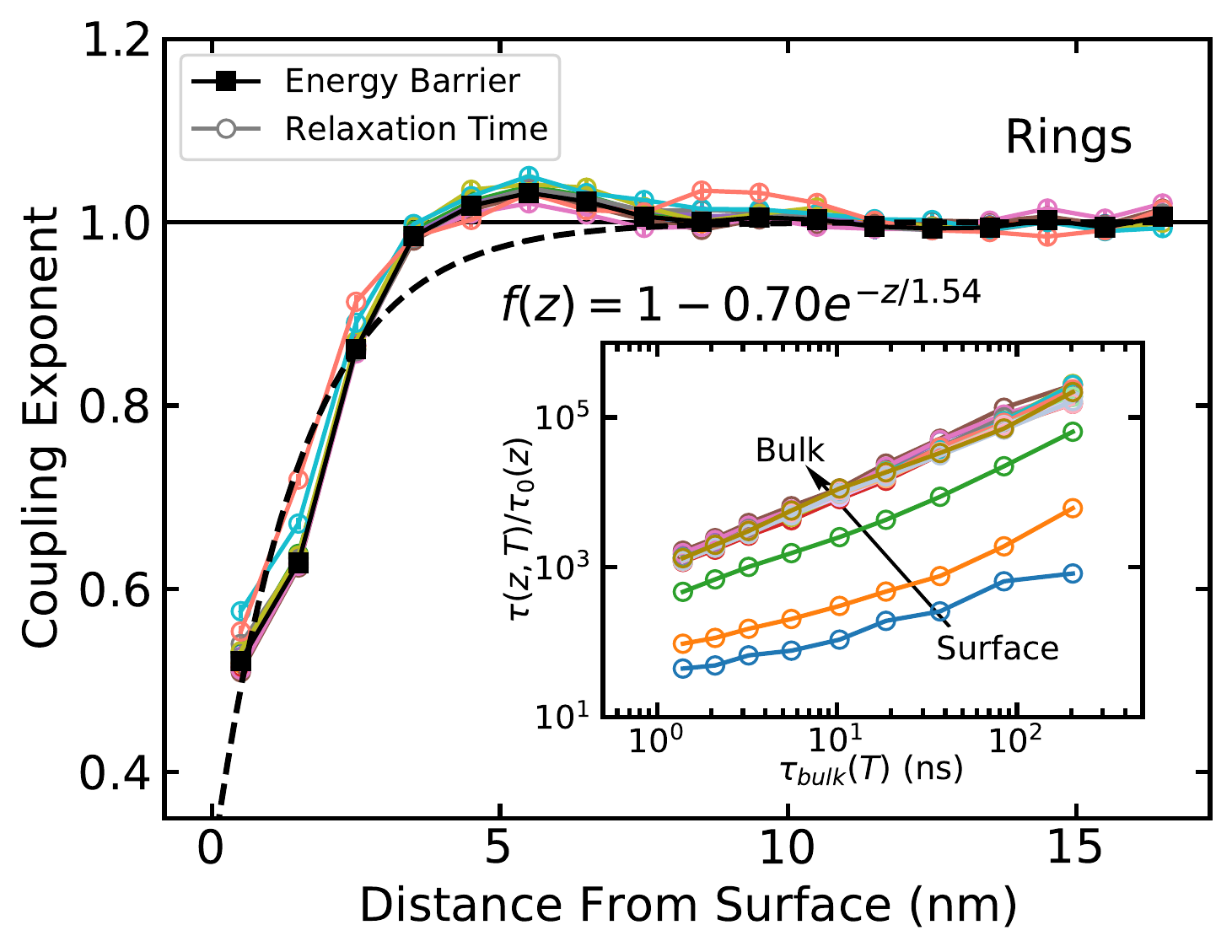}\\
\vspace{-0.05cm}
\includegraphics[width=0.9\columnwidth]{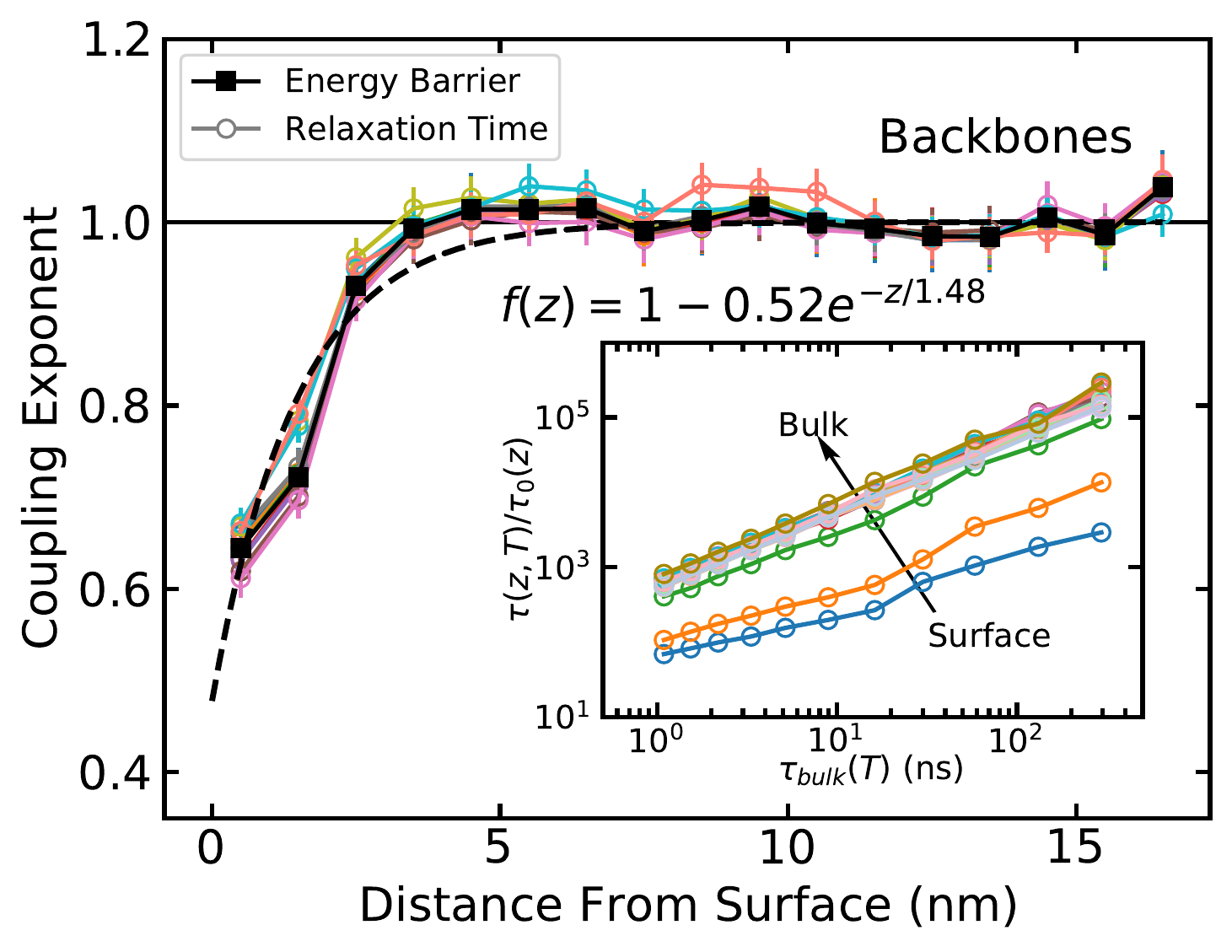}
\caption{Coupling exponent given by the logarithmic ratio of relaxation times $\log(\tau(T,z)/ \tau_0(z))/\log(\tau_\mathrm{bulk}(T)/\tau_0)$ (coloured, open) and ratio of activation barriers $\Delta E(z)/\Delta E_\infty$ (black, filled) from the \gls{vft} fits found in \Cref{fig:depth_vft,fig:depth_vft_bb} for rings (top) and backbones (bottom).  The insets show the relaxation times vs bulk relaxation time. The dashed line indicates an exponential fit, as calculated from the relaxation times. 
} 
\label{fig:tau}
\end{figure}

\begin{figure}[t]
\includegraphics[width=0.9\columnwidth]{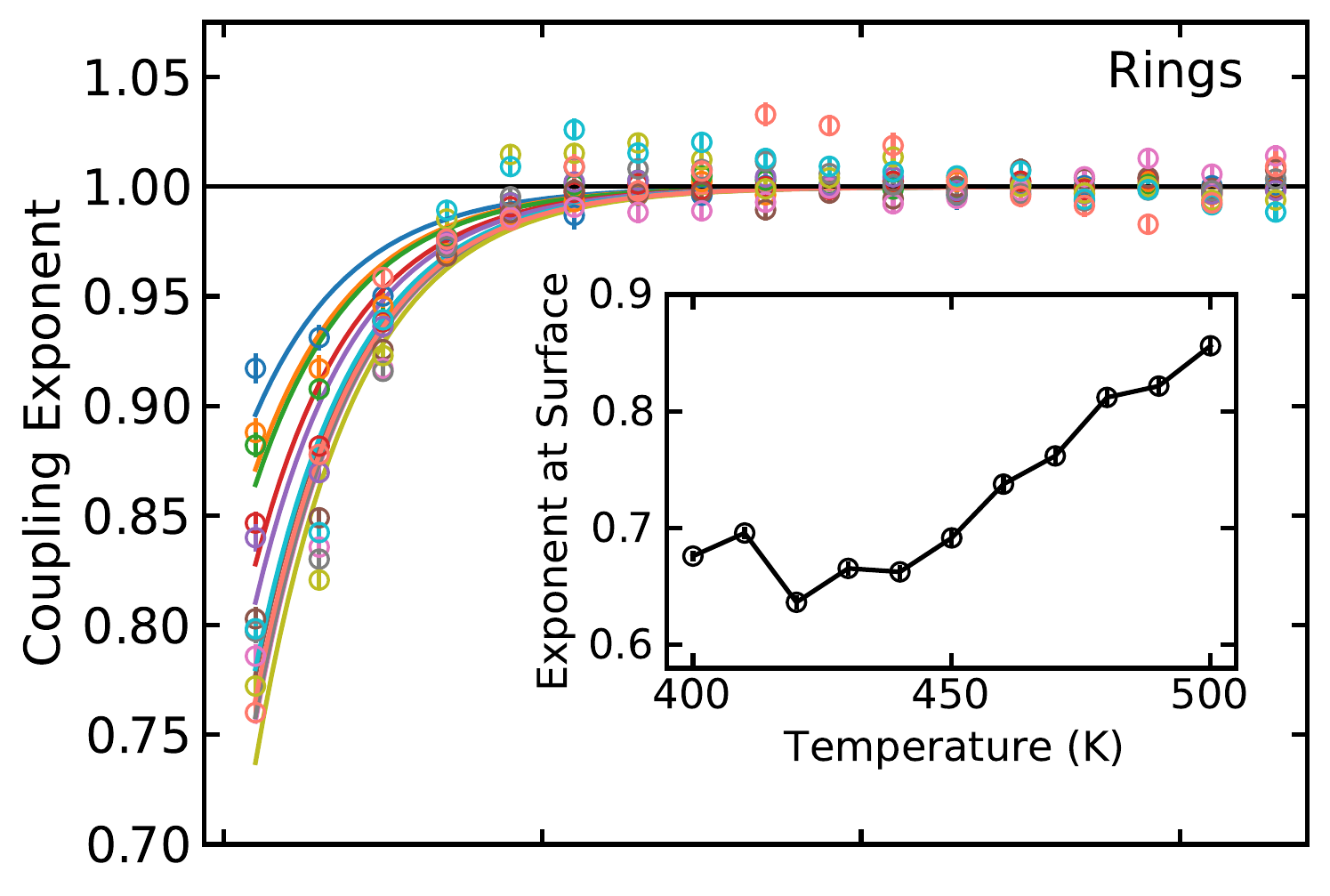}\\
\vspace{-0.2cm}
\includegraphics[width=0.9\columnwidth]{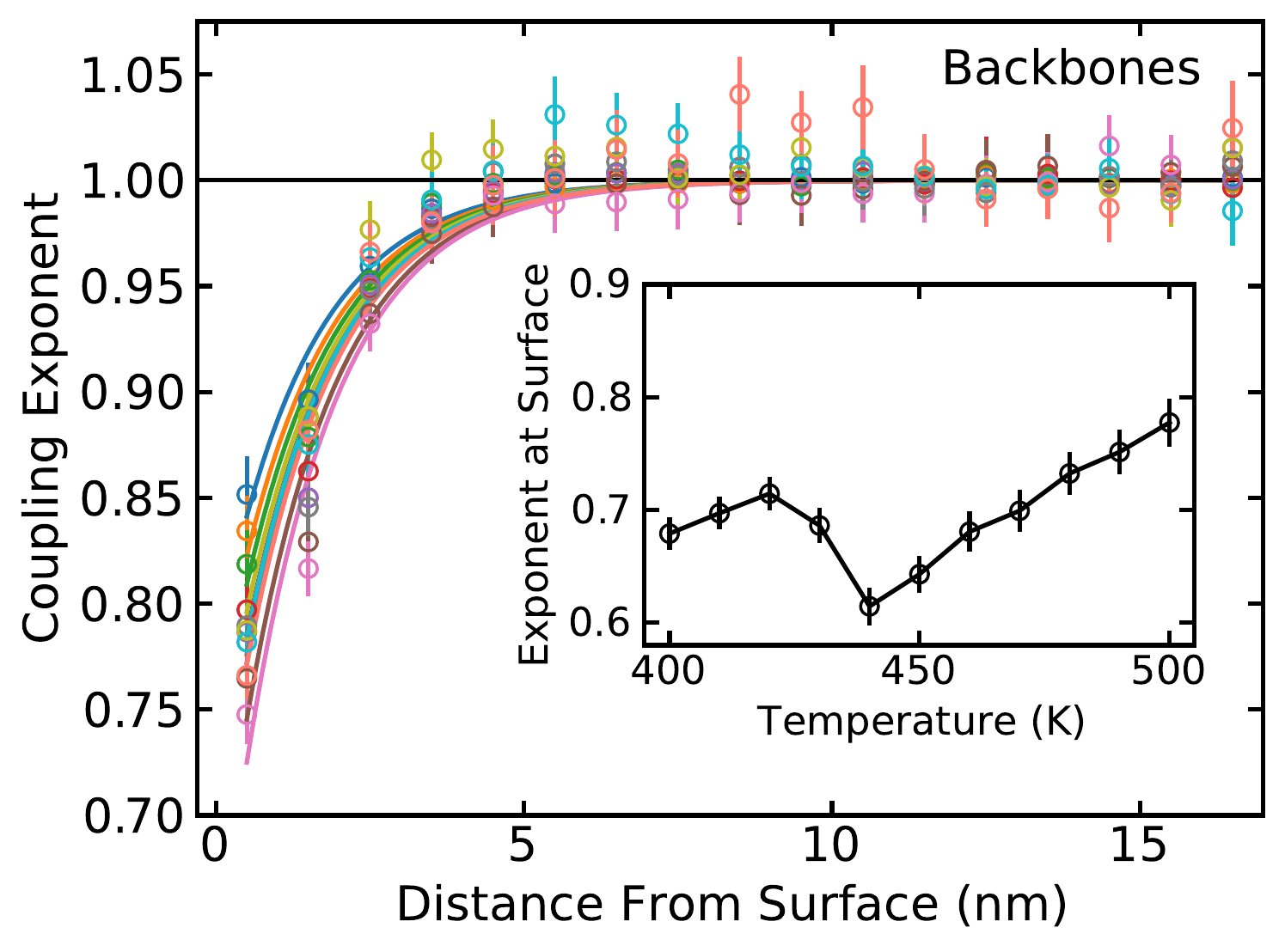}
\caption{Coupling exponent given by the logarithmic ratio of relaxation times $\log(\tau(T,z)/ \tau_0)/\log(\tau_\mathrm{bulk}(T)/\tau_0)$ for rings (top) and backbones (bottom) where the normalization $\tau_0$ is taken as the bulk value and independent of $z$. Fits are to the exponential form as in \Cref{fig:tau}, with the characteristic length scale $\xi$ fixed to the same as the corresponding fits in \Cref{fig:tau}. The inset shows the temperature dependence of the coupling exponent at $z=0$. }
\label{fig:tau_const}
\end{figure}

\begin{figure}[t]
\includegraphics[width=0.8\columnwidth]{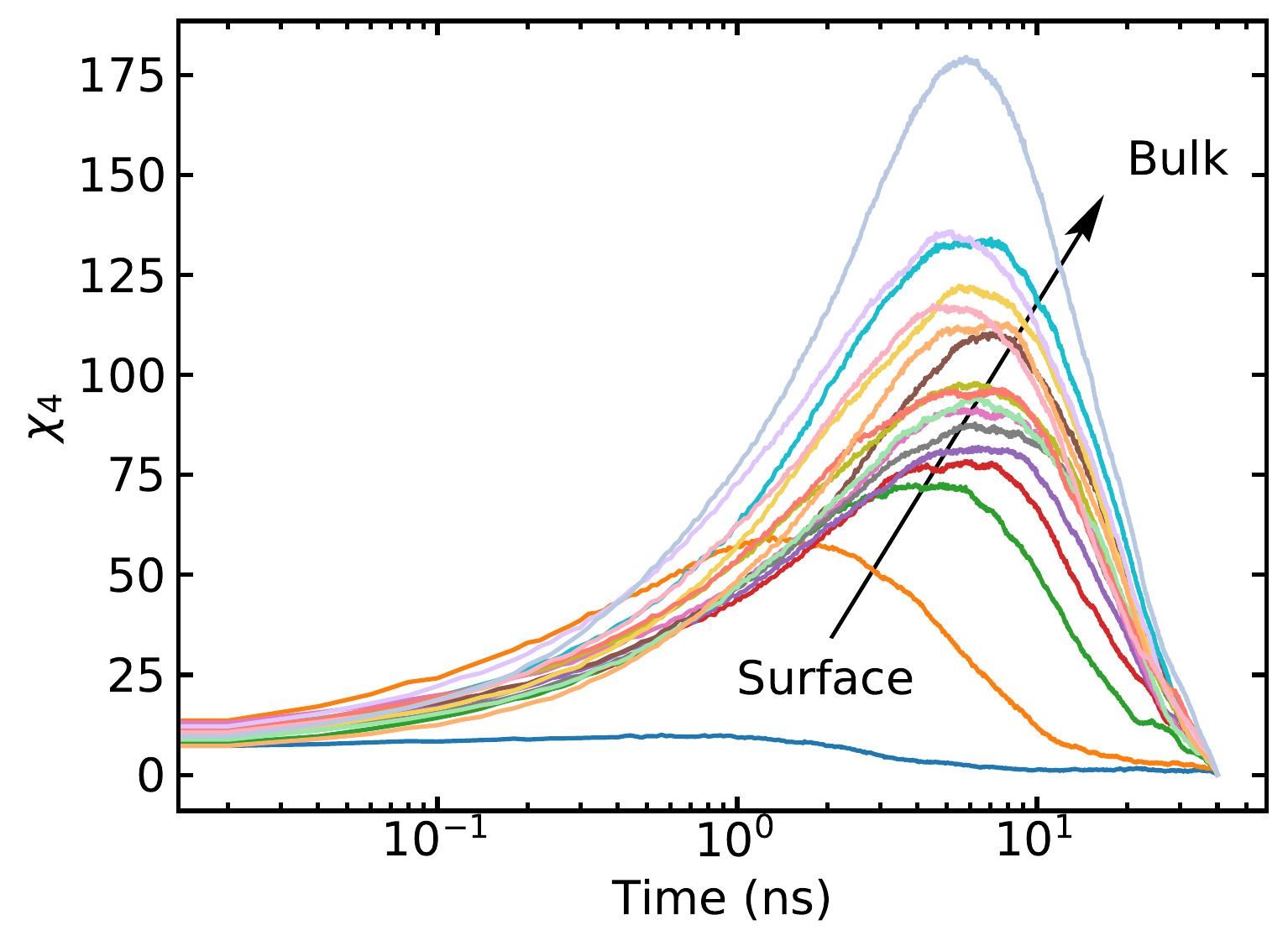}\\
\includegraphics[width=0.8\columnwidth]{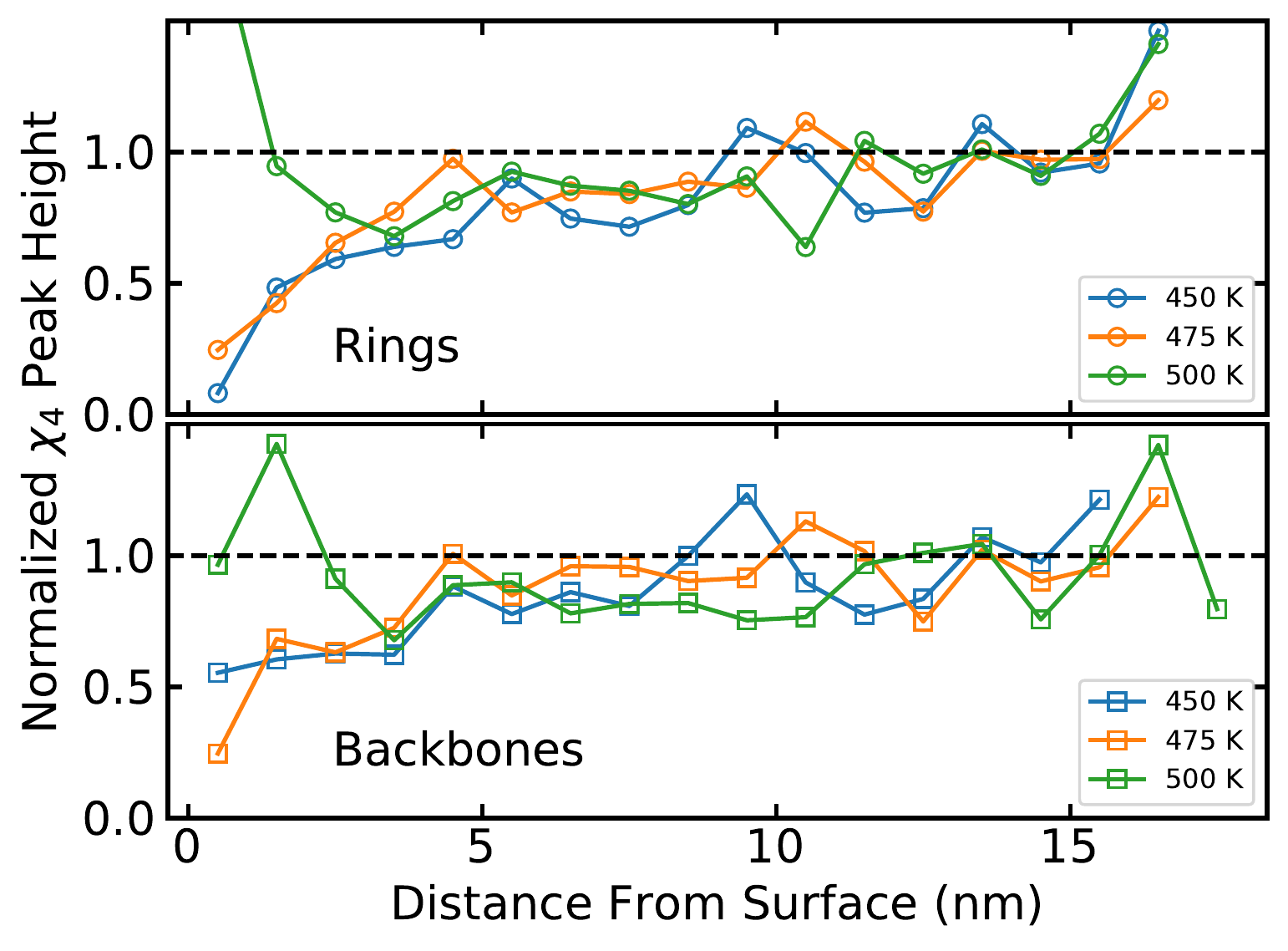}
\includegraphics[width=0.8\columnwidth]{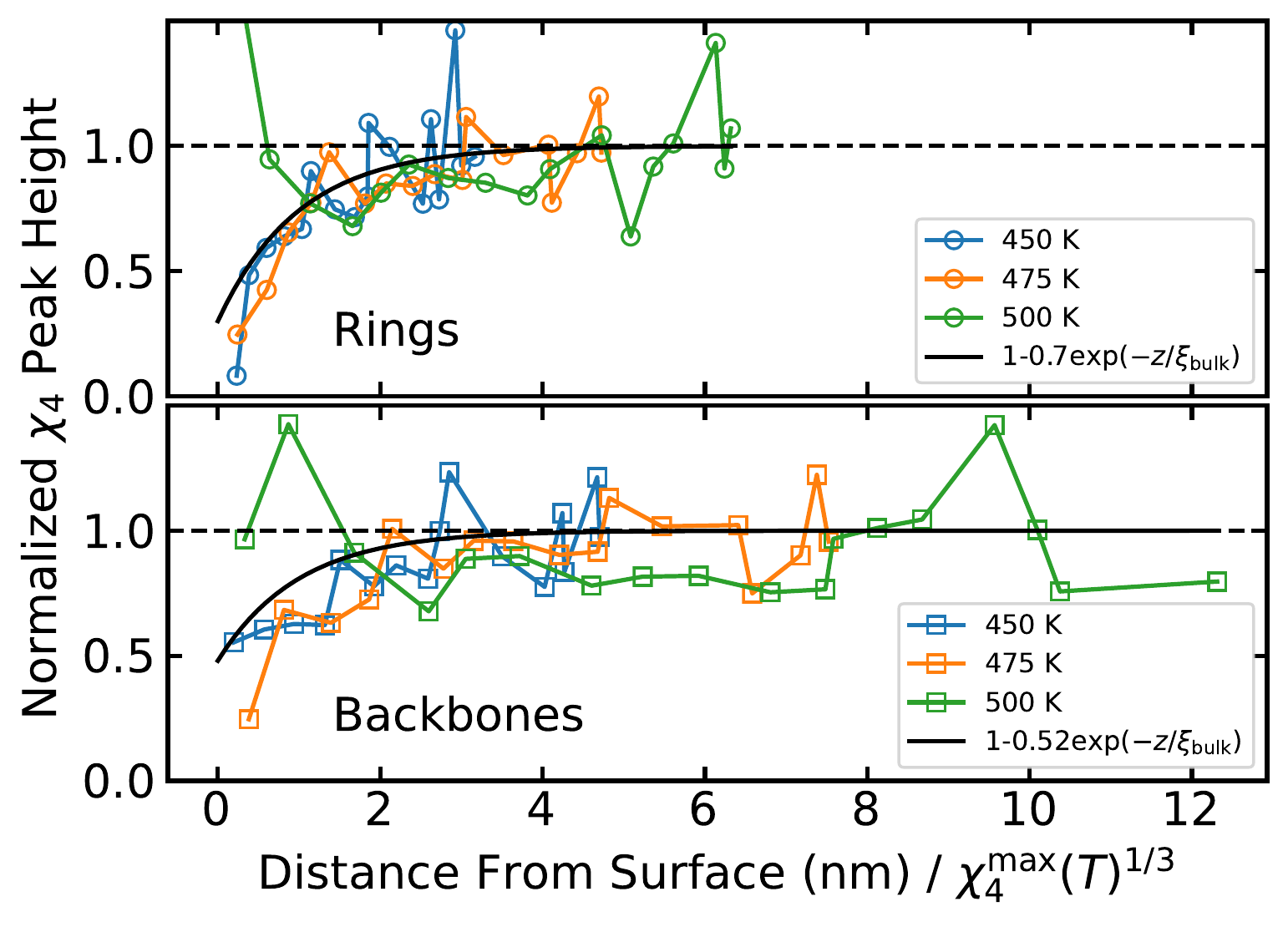}
\caption{\emph{(Top)} Dynamical susceptibility $\chi_4(T=\SI{450}{\K},z,t)$ for the phenyl rings. \emph{(Center)} Maximum of $\chi_4(T,z,t)$ vs. distance $z$ from the free surface normalized by $\chi_4^\mathrm{bulk}$, which is computed as an average of the 5 layers furthest away from the surface. \emph{(Bottom)} For comparison with the string model, we re-scale by $\chi_4^{1/3}(T)$, an estimate for $\xi_\mathrm{\revII{bulk}}(T)$. The surface point of the \SI{500}{\K} data (\num{\sim1.8}) was omitted for clarity. Black lines are produced from the fits in \Cref{fig:tau}.}
\label{fig:chi_raw}
\end{figure}

\Cref{fig:depth_vft} (top) and \Cref{fig:depth_vft_bb} (top) present the layer-resolved rotational relaxation time of the phenyl rings and backbones for temperatures $\SI{400}{\K} \le T \le \SI{500}{\K}$. The data represents an average over three independent trajectories of \SIrange{0.5}{1}{\micro\s} duration. The relaxation times decrease significantly in layers less than \SI{5}{\nm} from the free surface, converging quickly with increasing depth to a temperature-dependent bulk value. The depth dependence was fit with the phenomenological form $\log \tau = c_0 + c_1\text{erf}(z/z_0)$, as in \citet{Zhou2017}, but with shared $z_0$ across all temperatures, yielding dynamical length scales of $z_0=\SI{3.23\pm0.02}{\nm}$ and $z_0 = \SI{2.85\pm0.04}{\nm}$ for the rings and backbones, respectively. In our model, a freely varying $z_0$ produced length scales which did not vary appreciably in the temperature range studied. The length scales are comparable to that reported by \citet{Zhou2017} for the interfacial backbone relaxation and those found in bead-spring model simulations \cite{peter_thickness_2006,lang_interfacial_2013,shavit_physical_2014,sussman_disconnecting_2016}. 

The bottom panels replot the same data in an Arrhenius representation and also show \gls{vft} fits, where the activation barrier and pre-exponential factors were allowed to vary with $z$ but a single value of the \gls{vft} temperature $T_0$ was used over the entire data set. \revI{With a depth-dependent $T_0$, the model was overparametrized, resulting in artificial variability of the fit parameters.} The \gls{vft} fits were performed only for $T\ge \SI{410}{\K}$, since the behavior has been observed to cross over to a purely Arrhenius temperature dependence at temperatures near and below $T_\mathrm{g}$ \cite{lyulin2002correlated}. In the supercooled regime, the data follows the \gls{vft} form reasonably well. The inset shows that the logarithm of the pre-exponential timescale $\tau_0(z)$ is proportional to the activation barrier $\Delta E(z)$, a behavior often referred to as the Meyer-Neldel rule \cite{meyer1937relation,Yelon1992}. This can be interpreted as an entropy-enthalpy compensation effect and has also been observed in bead-spring models\cite{hanakata_interfacial_2014}. Results for the backbone motion mirror the behavior of the rings with longer relaxation times.

In order to test the validity of the explanations for near-surface relaxation, we plot in  \Cref{fig:tau} our data for \gls{ps} films in the form $\log(\tau(T,z)/ \tau_0(z))/\log(\tau_\mathrm{bulk}(T)/\tau_0)$ vs $z$ as suggested by \cref{eq:coupling_exp1}\revII{, where the value of $\tau_\mathrm{bulk}$ is obtained from the depth-independent film center}. This representation collapses curves for different $T$ onto a master curve and thus reveals a temperature independent coupling exponent $f(z)$. A fit to an exponential form suggests a short interfacial length scale $\xi\simeq$\SI{1.5}{\nm}. \revI{As a reference,  the average distance between two} \ce{CH2}-groups along the backbone is \SI{0.27}{\nm}. The insets show $\tau(z,T)/\tau_0(z)$ vs. $\tau_\mathrm{bulk}(T)$ in double-logarithmic form, so that the slope of the curves is the coupling exponent. Obtaining straight lines, we conclude that $f(z)$ depends only on $z$ and not on $T$. Our results are thus consistent with the proposal of \citet{diaz2018temperature} \revI{and the predictions of ECNLE theory \cite{phan2018dynamic,phan2019influence}} that the activation barrier at distance $z$ factorizes into distinct temperature- and depth-dependent parts,
\begin{equation}
    \Delta E(z)=f(z)\Delta E_\infty,
\end{equation}

As a further check of this relation, we can compare directly with the $z$-dependence of the activation barrier extracted from the \gls{vft} fits. The ratio $\Delta E(z)/\Delta E_\infty$ agrees strongly with the relaxation time data, and thus the form proposed in \cref{eq:coupling1}, as shown in \Cref{fig:tau}.
 
The above results clearly support the picture of a depth-dependent activation barrier driving the interfacial relaxation dynamics \cite{schweizer2019progress}. In \Cref{fig:tau_const} we examine the same data using a $z$-independent microscopic timescale $\tau_0$, which is assumed in the cooperative strings model and also in ECNLE theory. In this representation, the curves do not fully collapse but include a residual temperature dependence that is captured by exponential fits using the temperature independent length scales found in \Cref{fig:tau} (\SI{1.54}{\nm} or \SI{1.48}{\nm}), but allowing for temperature-dependent prefactors. As a result, the temperature dependence is carried by a variation of the coupling exponent $0.6<f(0)<0.9$ at the surface (see insets). This result is at variance with the cooperative string model that anticipates complete decoupling at the interface\cite{salez_cooperative_2015}, i.e. $f(0)\sim 0$. \revI{Our simulations cover the regime of weak to moderate supercooling, in which the assumed scaling form for the coupling exponent $f(z,T)=f(z/\xi_\mathrm{bulk}(T))$ might not yet apply. The absence of complete decoupling at the surface is however entirely compatible with ECNLE theory. \cite{schweizer2019progress}}

In order to probe the role of cooperativity more directly, we need a measure of the scale of dynamical heterogeneity. One possibility is to consider the layer-resolved variance of the autocorrelation function or four-point dynamical susceptibility \cite{lacevic2004,berthier2011}
\begin{equation}
\chi_4(T,z,t)=N_v(z)N_\tau(t)[\langle \bar{C}(z,t)^2\rangle-\langle \bar{C}(z,t)\rangle^2 ],    
\end{equation}
where $C$ is the \gls{acf} of an individual ring (backbone) vector as given by \cref{eq:autocorr}, the overbar denotes an average over $N_v(z)$ ring (backbone) vectors in a given layer, $\langle \rangle$ an average over 100 simulation instances and $N_\tau(t)$ the number of time slices used in the calculation of the  \gls{acf} for a given lag time $t$. $\chi_4(T,z,t)$ measures the fluctuations of the total molecular mobility as given by the backbone or phenyl ring dynamics. It can also be viewed as a (spatial) integral over a four-point correlation function that measures how the dynamics at locations ${\bf r}_1$ and ${\bf r}_2$ over a time interval $t=t_1-t_2$ are spatially correlated over a distance ${\bf r}={\bf r}_1-{\bf r}_2$. This function is shown for layers at different depths in the top panel of \Cref{fig:chi_raw} at temperature $T=\SI{450}{\K}$ as a function of time. All curves at different layers $z$ exhibit maxima at times that coincide with the layer-resolved relaxation times. The peak height can be interpreted as a correlation volume and thus proportional to the number of particles involved in a cooperative relaxation event. The middle panel plot this peak height $\chi_4^\mathrm{max}(T)$ normalized by the bulk value in the center of the film vs distance from the surface. While the data at \SI{500}{\K} does not exhibit any trend, we clearly see a reduction of cooperativity at the lower temperatures \SI{475}{\K} and \SI{450}{\K}. 

\revI{Reduced dynamical heterogeneity at the interface could for instance arise from a reduced collective barrier for activated processes as envisioned by ECNLE theory  \cite{xie2020collective,xie2020microscopic}. It could also accompany smaller cooperatively rearranging regions, i.e. shorter strings. A comparison with the cooperative string model is facilitated by rescaling $z$ by the bulk cooperativity length $\xi_\mathrm{bulk}(T)$, see \cref{eq:coupling2}}. In principle, this length scale could be extracted from the spatial decay of a four-point dynamical correlation function \cite{lacevic2004}. Here, we use instead a simple estimate  $\xi_\mathrm{bulk}(T)\propto \chi_4^\mathrm{max}(T)^{1/3}$, which is supported by simulations of a Lennard-Jones glass former \cite{flenner2011}. The bottom panel of \Cref{fig:chi_raw} shows that plotting the normalized $\chi_4^\mathrm{max}(T)$ data against $z/\chi_4^\mathrm{max}(T)^{1/3}$ leads to a reasonable collapse of our (limited) data set. The form of this master curve is overall consistent with the behavior of the coupling exponent $f(z)$ computed in \Cref{fig:tau} from the relaxation times (solid lines).

\section{Conclusions}
The relaxation times of backbone segments and phenyl rings at the surface of a freestanding \gls{ps} film were examined with \glsdesc{md}. The times $\tau$ are coupled to the bulk relaxation times via a power law relation with a temperature independent coupling exponent. These results extend previous bead-spring level simulations to a more detailed united atom model. The coupling exponent agrees well with the ratio of energy barriers extracted from \gls{vft}-fits, giving strong support to the notion that changes in the interfacial dynamics should be understood from interfacial changes in the activation free energy barrier. Moreover, the preexponential factors obey a Meyer-Neldel rule and thus exhibit considerable variation with depth below the free surface.

\revI{The near exponential variation of the coupling exponent with distance, the short and temperature independent characteristic length scale of \SI{\sim1.5}{\nm} and the absence of complete decoupling at the surface are all observations entirely consistent with the predictions of ECNLE theory. Our calculations also reveal a significant decrease of the dynamical four-point susceptibility near the surface, but cannot pinpoint the mechanism that is ultimately responsible for this behavior. }

If one accepts this measure of dynamical heterogeneity as a good characterization of cooperative motion, one can reconcile the coupling exponent with a normalized $\chi_4^\mathrm{max}(z,T)$ ratio. \revI{This does not prove, however, that varying string size controls the changes in  relaxation times.} Our results can be compared to a recent simulation study of the length $L$ of mobile strings in supported bead-spring polymer films \cite{zhang2019collective}. This work did not find any strong variation of $L$ across the film except very close to the free surface, and concluded that collective motion does not vary spatially in any strong manner. It must be noted, however, that the characteristic string time that maximizes the dynamical string length is shorter than the alpha-relaxation time that marks the peak of $\chi_4(t,T)$. For this reason, the $\chi_4^\mathrm{max}(z,T)$ parameter is more sensitive to slow particles as explained by \citet{starr2013relationship}. Future work could clarify the relationship between different measures of cooperativity in greater detail. 
 
\section{Acknowledgements}
J.R. thanks the Alexander von Humboldt Foundation for financial support, and D.F. acknowledges the support from a SBQMI QuEST fellowship.  This research was undertaken thanks, in part, to funding from the Canada First Research Excellence Fund, Quantum Materials and Future Technologies Program. 

\section{Data Availability}

Supporting data available upon reasonable request from the corresponding authors. 

\bibliography{glassy}
\end{document}